\documentclass[fleqn,twoside]{article}
\usepackage{espcrc2}
\usepackage{graphicx}
\usepackage[figuresright]{rotating}

\newcommand{\www}{\mbox{
\begin{minipage}{15pt}\includegraphics[width=14pt]{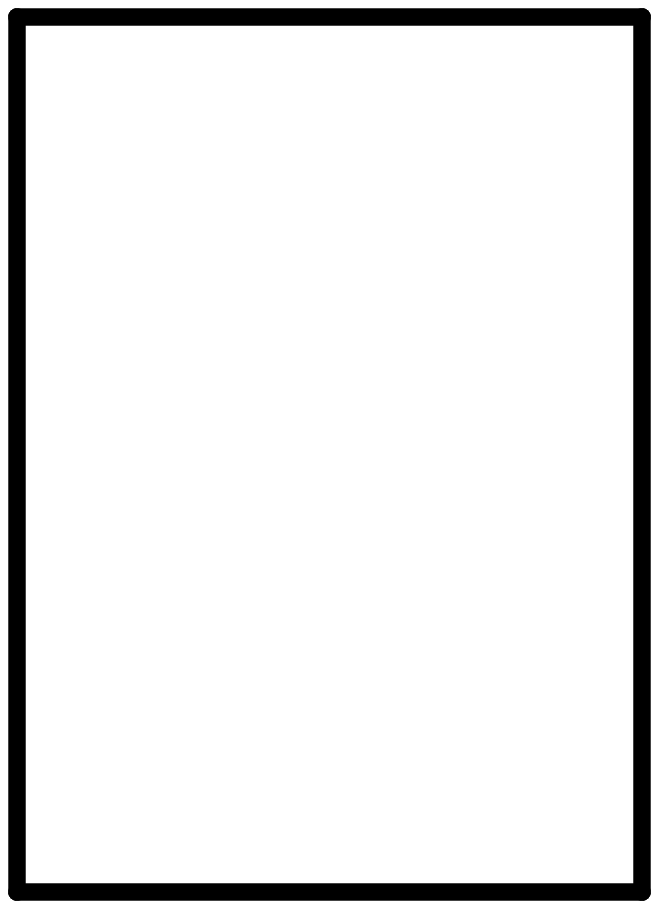}\end{minipage}}}
\newcommand{\wbbc}{\mbox{
\begin{minipage}{15pt}\includegraphics[width=14pt]{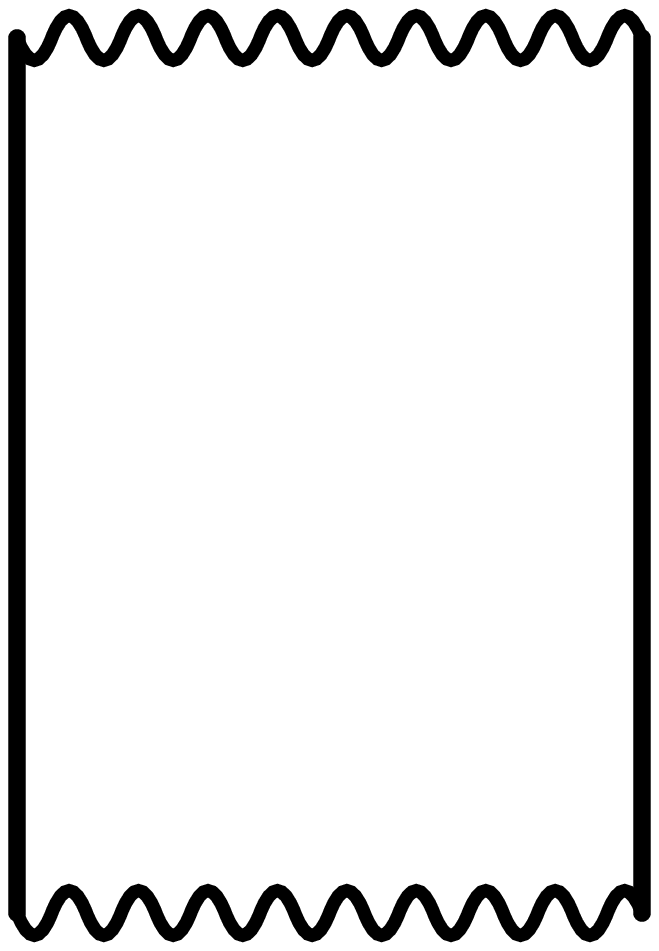}\end{minipage}}}
\newcommand{\wbbd}{\mbox{
\begin{minipage}{15pt}\includegraphics[width=14pt]{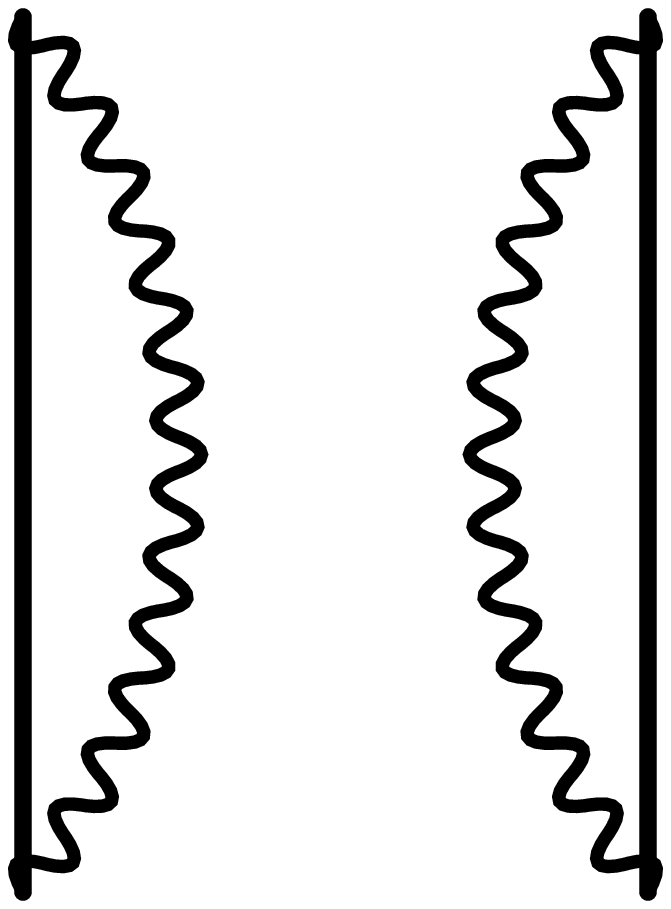}\end{minipage}}}
\newcommand{\wwb}{\mbox{
\begin{minipage}{15pt}\includegraphics[width=14pt]{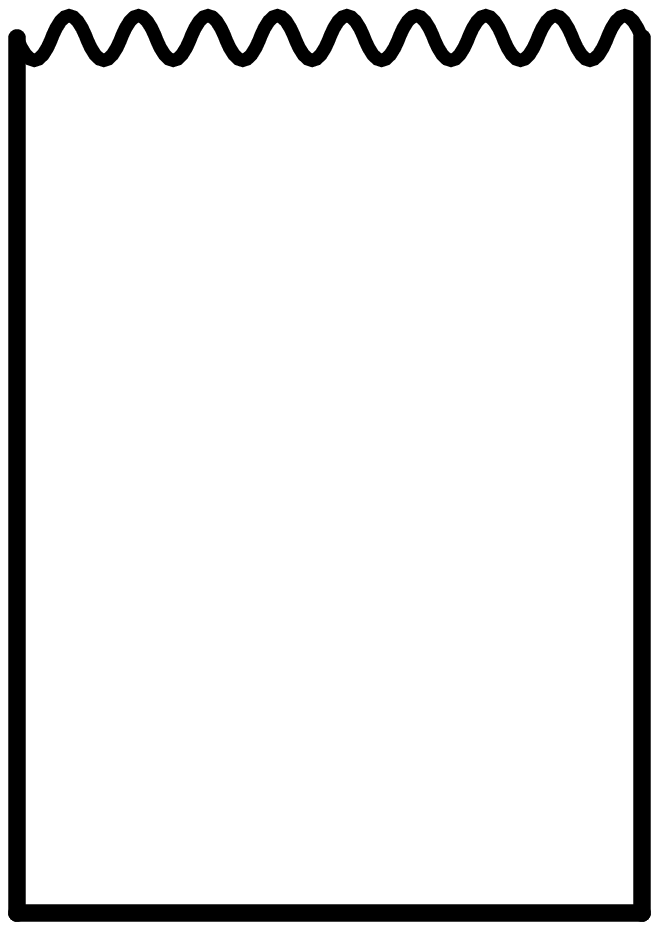}\end{minipage}}}
\newcommand{\wbw}{\mbox{
\begin{minipage}{15pt}\includegraphics[width=14pt]{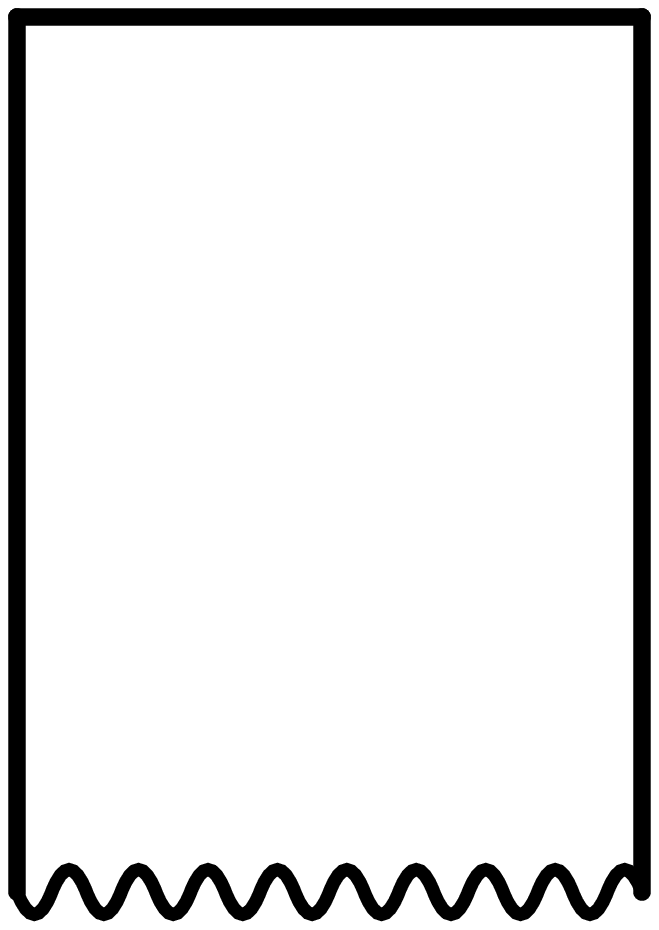}\end{minipage}}}
\newcommand{\ov}[1]{\overline{#1}}
\newcommand{\mpr}{{\frac{m_{\pi}}{m_{\rho}}}}

\newcommand{\AmS}{{\protect\the\textfont2
  A\kern-.1667em\lower.5ex\hbox{M}\kern-.125emS}}

\title{String breaking with dynamical Wilson fermions}

\author{G.\ S.\ Bali\address[GL]{Department of Physics \& Astronomy, 
        The University of Glasgow, Glasgow G12 8QQ, Scotland},
        Th.\ D\"ussel\address[JU]{Zentralinstitut f\"ur Angewandte Mathematik,
        Forschungszentrum J\"ulich, D-52425 J\"ulich, Germany}
        Th.\ Lippert\addressmark[JU]\thanks{th.lippert@fz-juelich.de},
        H.\ Neff\address[BU]{Center for Computational Science, 
        Boston University, 3 Cummington St., Boston MA02215, USA},	
        Z.\ Prka\c cin\address[WU]{Fachbereich Physik, Bergische 
        Universit\"at Wuppertal, Gau\ss{}stra\ss{}e, D-42097 Wuppertal, Germany},
        K.\ Schilling\addressmark[WU]}
       
\begin{document}

\begin{abstract}
We present results of our ongoing determination of string breaking in
full QCD with $N_f=2$ Wilson fermions.  Our investigation of the
fission of the static quark-antiquark string into a static-light
meson-antimeson system is based on dynamical configurations of size
$24^3 \times 40$ produced by the T$\chi L$ collaboration.  Combining
various optimization methods we determine the matrix elements of the
two-by-two system with so far unprecedented accuracy.  The all-to-all
light quark propagators occurring in the transition element are
computed from eigenmodes of the Hermitian Wilson-Dirac matrix
complemented by stochastic estimates in the orthogonal subspace.  We
observe a clear signature for level-splitting between ground state and
excited potential. Thus, for the first time, string breaking induced
by sea quarks is observed in a simulation of 4-dimensional
lattice-QCD.
\vspace{1pc}
\end{abstract}

\maketitle

\section{INTRODUCTION}
The fission of the flux string between two heavy quarks, is a key
problem of lattice quantum chromodynamics (QCD). While it is
fundamental to a proper understanding of QCD, after nearly 25 years
of lattice QCD we are still lacking a convincing demonstration of
string breaking through dynamical quarks \cite{Schilling:1999mv}.

Of course, extremely precise studies of the potential have been
carried out, however, even at separations $r=|R|a$ much
larger than the anticipated breaking distance of about 1 fm, still a
linearly rising potential is found. This phenomenon is attributed to
the insufficient overlap of the Wilson loop with the final state at
large $r$.

As first suggested back in 1992 by C.\ Michael, progress in the
conceptual understanding of string breaking on the lattice could be
achieved: in Ref.~\cite{Knechtli:2000df}, QCD was mimicked by the
the SU(2)-Higgs model using a two channel setting.  The
authors could confirm the interpretation of string-breaking as a {\it
dynamical} level-crossing phenomenon between string-type and
meson-type states. They observed a pronounced gap of the energy levels
at the location of the crossing.

A main hindrance of a straightforward application of this procedure to
4-dimensional QCD is the computation of {\it all-to-all} propagators
to achieve a proper signal for the heavy-light contributions.  The
elements of the $2\times 2$ mixing matrix---three of which are
composites of Wilson lines and fermion propagators--- must be averaged
over all spatial lattice sites and dimensions.  This requires the
computation of the inverse of the Hermitian Wilson-Dirac matrix,
$Q^{-1}$. As the dimension of $Q$ is about ${\cal O}(10^{7})$, the
full inverse cannot be represented even on high-end systems.

In our investigation, we employ various ``smearings'' like APE
smearing, Wuppertal smearing, fat temporal links and a new hopping
parameter acceleration. The latter is shown to enhance the low
eigenmode dominance. Furthermore, with the truncated eigenmode
approximation (TEA) \cite{Neff:2001zr}, rendered exact by stochastic
estimates on the orthogonal subspace, all-to-all propagators can be
computed efficiently.

\section{MODELLING THE TRANSITION}

Close to the location of QCD string breaking, we deal with a mixture of two 
states, the initial heavy $Q\bar Q$-pair at small separation $r$,
represented by the standard Wilson-Loop $W(R,T)$, and a final state at
large separation $r$, which is given by two {\sl heavy-light}
mesons.  The interesting physics, {\it i.e.}\ the fission of the
flux-tube, takes place at intermediate $r$.

In the two-channel approach the correlator is represented as
the following $2\times 2$-matrix
\begin{eqnarray}
C(t)&=&e^{-2m_Qt}\left(\begin{array}{rr}
\www&\sqrt{n_f}\wwb\\
\sqrt{n_f}\wbw&\wbbd-n_f\wbbc\end{array}\right)
\nonumber\\
& = &e^{-2m_Qt}\left( \begin{array}{cc} C_{1,1} & C_{1,2} \\ 
C_{2,1} & C_{2,2} \end{array} \right)
\label{SBCorrelator}
\end{eqnarray}
$C_{1,1}$ is the correlation function of the static $Q\bar Q$ pair, namely
the Wilson loop, $C_{2,2}$ is the correlator of the heavy-light pair,
$B\ov{B}$, and $C_{1,2}$ the correlator of the transition matrix
element. The light quark propagators in the generalized Wilson loops
are represented by wavy lines.

The exponential decay of the correlator is found by diagonalization of
the real $2\times 2$ transfer matrix, with the lower eigenvalue being
the ground state energy of the system.

As demonstrated in Ref.~\cite{Knechtli:2000df}, characteristic
signatures are required in order to ``prove'' string breaking:
\begin{enumerate}
\item\vspace*{-.2cm}
The energy levels show crossing with the asymptotic ground state
reaching the same mass as the system of two separated heavy-light
mesons.
\item\vspace*{-.2cm}
At the location of crossing, dynamical string breaking manifests
itself in a pronounced energy level-splitting.
\item\vspace*{-.2cm}
The transition matrix element $C_{12}$ will stay finite for
$a\rightarrow 0$.
\end{enumerate}

\section{SETTING}

The basis for our investigation of string breaking are the $N_f=2$
dynamical configurations produced by the SESAM and T$\chi$L groups.
SESAM generated 8 ensembles of about 200 decorrelated configurations
each on $16^3\times 32$ lattices at two different $\beta$-values, 5.6
($\kappa= 0.156, 0.1565, 0.157, 0.1575, 0.1580$, $0.56<\mpr<0.83$) and
5.5 ($\kappa=0.158, 0.159, 0.1596, 0.1600$, $0.67<\mpr<0.85$). These
ensembles have been analyzed for  string breaking in
order to determine a $\beta-\kappa$ combination with small enough sea
quark mass still being unaffected by finite-size effects. We have
chosen the pair $\beta=5.6$, $\kappa=0.1575$.

With a Sommer scale of $r_0=5.892(27)a$ at $r=0.5$ fm the lattice
spacing is $a=0.085$ fm or $a^{-1}= 2.32$ GeV.  $\mpr=0.704(5)$
suggests a degenerate quark mass close to the strange quark mass.  For
the $16^3\times 32$ system, both the pion window of $La \approx 4
m_{\pi}^{-1}$ and the string breaking window with $\sqrt{2}L/2r_{SB}
\approx .96 <1$ and $\sqrt{3}L/2r_B \approx 1.2 \ge 1$ are not 
large enough to comfortably accommodate the string breaking experiment. We
chose to use our T$\chi$L configurations generated on $24^3\times 40$
lattices. Here the windows are $La \approx 6 m_{\pi}^{-1}$ and
$\sqrt{2}L/2r_{SB} \approx 1.2 >1$, $\sqrt{3}L/2r_B \approx 1.4 \ge
1$. 183 decorrelated configurations were analyzed for Wilson loops, 20
decorrelated configurations were analyzed for fermion lines.

\section{OVERLAP ENHANCEMENTS}

We use a set of mutually optimized methods to improve the ground 
state overlap which reduce the statistical error by a least a factor of 5:\\[6pt]
\noindent{\bf APE smearing.}
APE smearing is carried out on the gauge fields with 
$\alpha= 2.0$ and $N=50$:
\begin{eqnarray}
\vspace*{-.1cm}
U^\prime_i(\vec x)& =& \mathcal{P}_{SU(3)} \Big[U_i(\vec x)\\
& +&\frac{1}{\alpha} \sum_{j\neq i} \Big(U_i(\vec x)U_j(\vec x+ \hat
i)U_i^\dagger(\vec x+\hat j)\nonumber\\ &+& U_i^\dagger(\vec x-\hat
i)U_j(\vec x- \hat i)U_i(\vec x-\hat i +\hat j)\Big)\Big]\nonumber
\end{eqnarray}
\noindent{\bf Wuppertal smearing} with $\delta=4$ and linear combinations of $N_W=20,40,50$:
\vspace*{-.1cm}
\begin{equation}
\phi^{(n+1)}_x=\frac{1}{1+6\delta}\Bigg(\phi^{(n)}_x+
\delta\sum_{j=\pm 1}^{\pm 3}U_{x,j}\phi^{(n)}_{x+a\hat{\jmath}}\Bigg)
\end{equation}
\noindent{\bf Fat temporal links (HYP)} \`a la Hasenfratz and Knechtli
can lead to an exponential improvement of the signal quality.\\
\noindent{\bf Hopping parameter acceleration (HPA).} The  $k$th power
of the Wilson hopping term, $(\kappa D)^k$, with  $M=1-\kappa D$, connects
sites at distance $k$ and less.  Multiplying
$M^{-1}=1+(\kappa D)+(\kappa D)^2+(\kappa D)^3\dots$ by $(\kappa D)^k$
leads to $M^{-1}(\kappa D)^k=M^{-1}-\sum_{i=1}^{k-1} (\kappa D)^i$.
Thus, at distance $k$, the first $k-1$ terms do not contribute, and
the low eigenmode dominance is enhanced, since with
$M|\psi\rangle=\lambda|\psi\rangle$ we have $M^{-1}(\kappa
D)^k|\psi\rangle=\frac{(1-\lambda)^k}{\lambda}|\psi\rangle$.

\begin{figure}[!htb]
\vspace*{-.4cm}
\centerline{\includegraphics[width=.7\columnwidth]{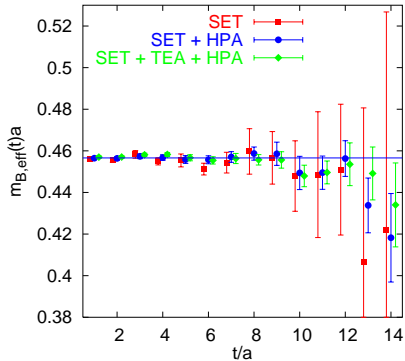}}
\vspace*{-1cm}
\caption{Improvements by HPA and TEA.}
\vspace{-.9cm}\label{fig:SET}
\end{figure}

Furthermore, we improve $Q^{-1}$ by the {\bf Truncated Eigenmode
Approximation (TEA)} \cite{Neff:2001zr}.  TEA is made exact by
means of stochastic estimates (SET) in the orthogonal subspace.  In
Fig.~\ref{fig:SET} we compare the standard SET-method (no mode
computed via TEA) on the smeared gauge fields with the improvement
through HPA and TEA (200 modes).

\section{RESULTS}
\begin{figure}[ht]
\vspace*{-.1cm}
\includegraphics[width=\columnwidth]{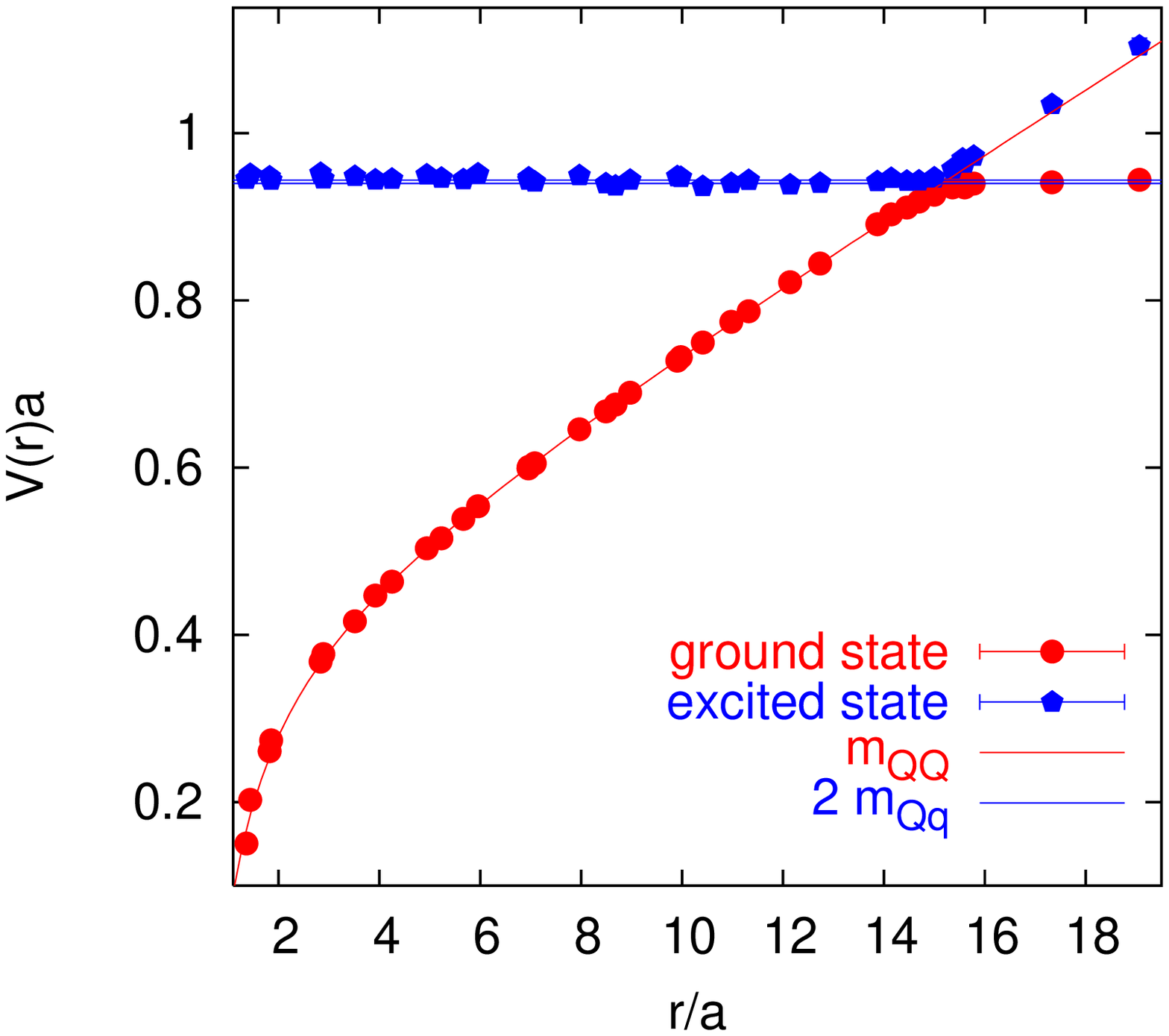}
\vspace*{-.3cm}
\includegraphics[width=\columnwidth]{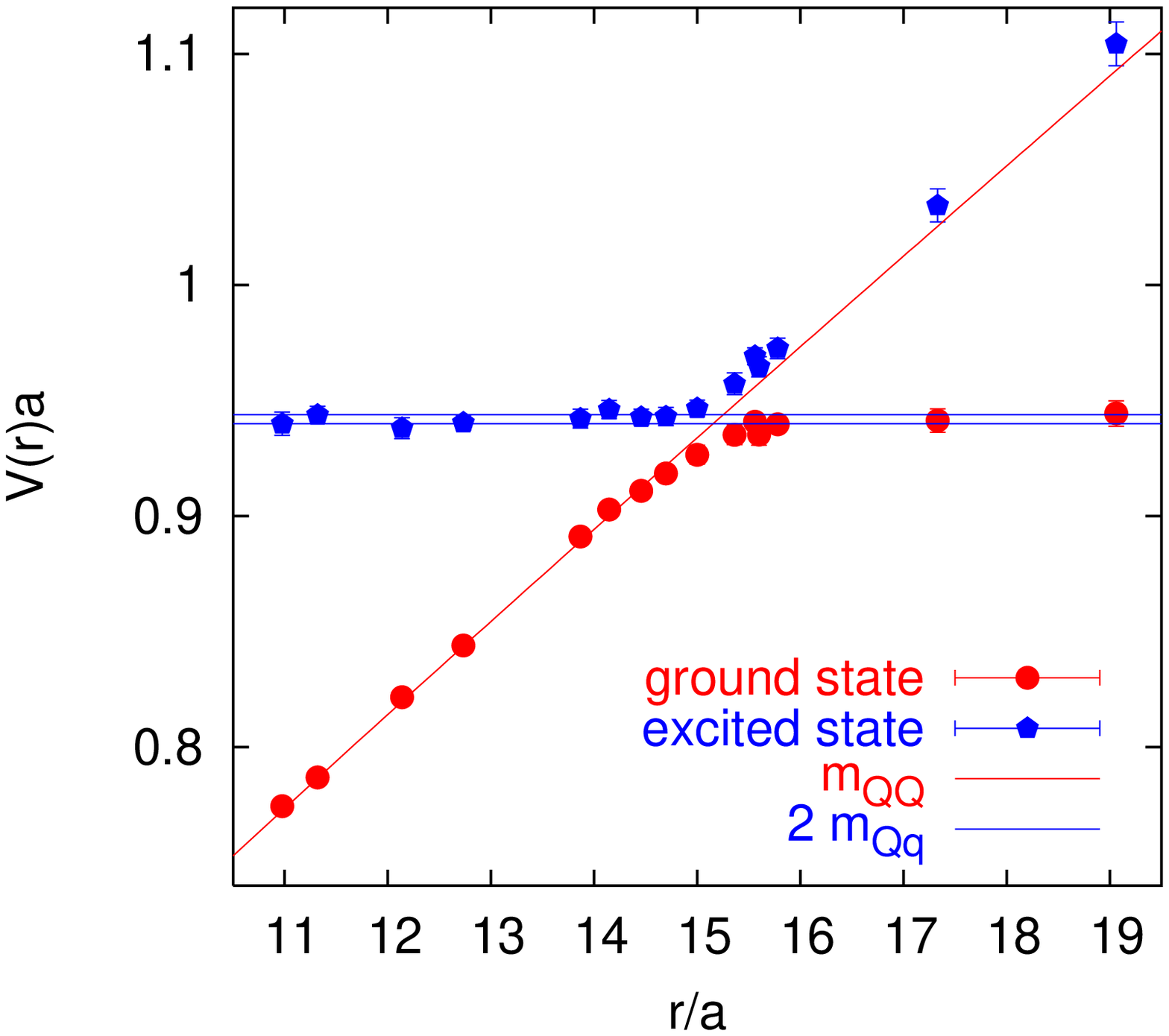}
\vspace*{-1cm}
\caption{Appearance of a level-gap.}
\vspace*{-.9cm}
\label{fig:SBA}
\end{figure}

Fig.~\ref{fig:SBA} shows intermediate results of our ongoing
evaluation of T$\chi$L configurations that have been obtained with a
preliminary form of the fat link static action.  Hence the asymptotic
ground state is just slightly higher than twice the heavy-light mass.  The
error bars of the zoomed plot are smaller than the symbols.
Level-splitting is achieved at the crossing point for the first time
in 4-dimensional QCD with dynamical fermions, demonstrating string
breaking.

\vspace*{.1cm}\noindent{\bf Acknowledgments.}  The computations were performed on
the IBM Regatta p690$+$ of the ZAM, FZ-J\"ulich, and ALiCE, Wuppertal
University.  We thank the John von Neumann Institute for Computing for
granting computer time.  This work is supported by the Deutsche
Forschungsgemeinschaft under contract Li701/4-1 and by the EC Hadron
Physics I3 Contract No.\ RII3-CT-2004-506078.  G.B.\ is supported by a
PPARC Advanced Fellowship (grant PPA/A/S/2000/00271) as well as by
PPARC grant PPA/G/0/2002/0463.\\[-.7cm]

\end{document}